\documentclass[%
 reprint,
superscriptaddress,
 amsmath,amssymb,
 aps,
prb,
floatfix,
]{revtex4-2}

\usepackage{graphicx}
\usepackage{dcolumn}
\usepackage{bm}
\usepackage[version=3]{mhchem} 
\usepackage[T1]{fontenc}
\usepackage{siunitx}
\usepackage[T1]{fontenc}
\usepackage{siunitx}
\usepackage[export]{adjustbox}
\usepackage{hyperref}
\usepackage{gensymb}

\begin{document}


\title[Femtosecond laser-written nano-ablations containing bright antibunched emitters on gallium nitride]{Femtosecond laser-written nano-ablations containing bright antibunched emitters on gallium nitride}
\author{Yanzhao Guo}
\email{GuoY65@cardiff.ac.uk}
\affiliation{School of Engineering, Cardiff University, Queen’s Buildings, The Parade, Cardiff, CF24 3AA, United Kingdom}
 \affiliation{Translational Research Hub, Cardiff University, Maindy Road, Cardiff, CF24 4HQ, United Kingdom}
 \author{Giulio Coccia}
\affiliation{Department of Physics, Politecnico di Milano, Piazza Leonardo da Vinci, 32, 20133 Milano, Italy}
\affiliation{Institute for Photonics and Nanotechnologies (CNR-IFN), Piazza Leonardo da Vinci, 32, 20133 Milano, Italy}
\author{Vibhav Bharadwaj}
\affiliation{Institute for Photonics and Nanotechnologies (CNR-IFN), Piazza Leonardo da Vinci, 32, 20133 Milano, Italy}
\affiliation{Department of Physics, Indian Institute of Technology, Guwahati, Assam, India 781039}
\author{Reina Yoshizaki }
\affiliation{Department of Mechanical Engineering and the 
Research into Artifacts, Center for Engineering (RACE), University of Tokyo}
\author{ Katie M. Eggleton}
\affiliation{School of Engineering, Cardiff University, Queen’s Buildings, The Parade, Cardiff, CF24 3AA, United Kingdom}
\affiliation{Translational Research Hub, Cardiff University, Maindy Road, Cardiff, CF24 4HQ, United Kingdom}
 \author{John P. Hadden}
\affiliation{School of Engineering, Cardiff University, Queen’s Buildings, The Parade, Cardiff, CF24 3AA, United Kingdom}
\affiliation{Translational Research Hub, Cardiff University, Maindy Road, Cardiff, CF24 4HQ, United Kingdom}
\author{Shane M. Eaton}
\affiliation{Department of Physics, Politecnico di Milano, Piazza Leonardo da Vinci, 32, 20133 Milano, Italy}
\affiliation{Institute for Photonics and Nanotechnologies (CNR-IFN), Piazza Leonardo da Vinci, 32, 20133 Milano, Italy}
\author{Anthony J. Bennett}
\email{BennettA19@cardiff.ac.uk}
\affiliation{School of Engineering, Cardiff University, Queen’s Buildings, The Parade, Cardiff, CF24 3AA, United Kingdom}
\affiliation{Translational Research Hub, Cardiff University, Maindy Road, Cardiff, CF24 4HQ, United Kingdom}

\date{\today}

\begin{abstract}
Femtosecond laser-writing offers distinct capabilities for fabrication, including three-dimensional, multi-material, and sub-diffraction-limited patterning. In particular, demonstrations of laser-written quantum emitters and photonic devices with superior optical properties have attracted attention. Recently, gallium nitride (GaN) has been reported to host quantum emitters with narrow and bright zero-phonon photoluminescence from ultraviolet to telecom ranges. However, emitters formed during epitaxy are randomly positioned, and until now, it has not been possible to fabricate quantum emitters in ordered arrays. In this paper, we employ femtosecond laser writing to create nano-ablations with sub-diffraction-limited diameter, and use rapid thermal annealing to activate co-located stable emitters. The emitters show \SI{}{MHz} antibunched emission with a sharp spectral peak at room temperature. Our study not only presents an efficient approach to laser-written nanofabrication on GaN but also offers a promising pathway for the deterministic creation of quantum emitters in GaN, shedding light on the underlying mechanisms involved.
\end{abstract}

\maketitle

\section{\label{Introduction} Introduction}

In the last decade, quantum emitters (QEs) in gallium nitride (GaN) have been investigated as a practical source of quantum light at room temperature\cite{Castelletto2024GalliumTechnologies}. These QEs display low multi-photon emission probability\cite{Berhane2017BrightNitride}, comparatively high Debye-Waller factor\cite{Luo2024RoomGaN}, high continuous wave (CW) photon detection rate\cite{Bishop2022EnhancedLens} and telecom range emission\cite{Zhou2018RoomRange}. Recently, resonant optical excitation\cite{Kianinia2018ResonantNitride} and optically detected magnetic resonance have been observed in GaN QEs\cite{Luo2024RoomGaN,PhysRevLett.134.083602}, pointing to potential applications in quantum sensing and quantum networks. However, the reliable fabrication of QEs in GaN remains a challenge. This greatly hinders high-yield coupling to photonic structures that manipulate the local density of optical states and enhance photon collection\cite{Elshaari2020HybridCircuits}, which is crucial for the realization of scalable integrated quantum photonics. 

Previous approaches to creating these QEs, such as ion implantation and doping during growth\cite{Peng2021FormationGaN,Nguyen2019EffectsNitride,Nguyen2021SitePolarity,Berhane2017BrightNitride,Luo2024RoomGaN}, create QEs with random positions and residual lattice damage\cite{Eaton2019QuantumIrradiation,GUO202547} which one may expect to degrade the spin and optical coherence properties. Recently, femtosecond laser-writing fabrication has attracted intense attention for its precise and highly controlled creation of QEs in semiconductors, such as diamond\cite{Chen2017LaserDiamond,Chen2019LaserYield,cheng2025laser}, silicon carbide \cite{Chen2019LaserCarbide}, hexagonal boron nitride\cite{Gao2021FemtosecondNitride,Yang2023LaserTechnologies}, and aluminum nitride\cite{Wang2023QuantumLaser}. Moreover, femtosecond laser-writing is also used as a promising method to fabricate 3D photonic circuits from nanoscale to microscale\cite{Eaton2019QuantumIrradiation,10.1063/5.0160067,GUO202547}. Direct laser-writing offers a maskless fabrication process below the optical diffraction limit. Under optimal fabrication conditions, laser-written waveguide-integrated QEs in diamond feature spin coherence properties comparable to their native counterparts in the host material\cite{Guo2024Laser-writtenDiamond,doi:10.1021/acs.nanolett.5c00148}. 

In this paper, we report femtosecond laser-writing and subsequent annealing to fabricate an array of emitters in GaN. Because the bright spots observed in the PL scan are in a regular square array, we can be certain that the laser writing induces them, and not intrinsic to the semiconductor. As we will show, bright emission is correlated in position with the nano-ablated holes on the sample surface. Laser-written nano-ablations in GaN are mapped by atomic force microscopy (AFM) and confocal photoluminescence (PL) mapping. Compared to previously studies which reported broadband PL emission\cite{Saleem:18,Castelletto2020ColorSemiconductors}, the laser-written structures in this paper exhibit a sharp emission peak in their spectrum at room temperature, antibunching in their photon emission correlation spectra (PECS), shelving behavior in time-resolved PL (TRPL) measurements, and few \SI{}{MHz} saturation PL rates under CW excitation. These features are consistent with the creation of a small number of QEs in a sub-diffraction-limited spot. Our study provides a promising route to scalable, integrated, on-chip quantum technologies based on GaN via femtosecond laser writing.  

\section{\label{Results and discussion} Results and discussion}
\subsection{\label{Femtosecond laser-writing fabrication on GaN}Femtosecond laser-writing fabrication on GaN}

\begin{figure*}[ht]
    \centering
    \includegraphics[width=\textwidth]{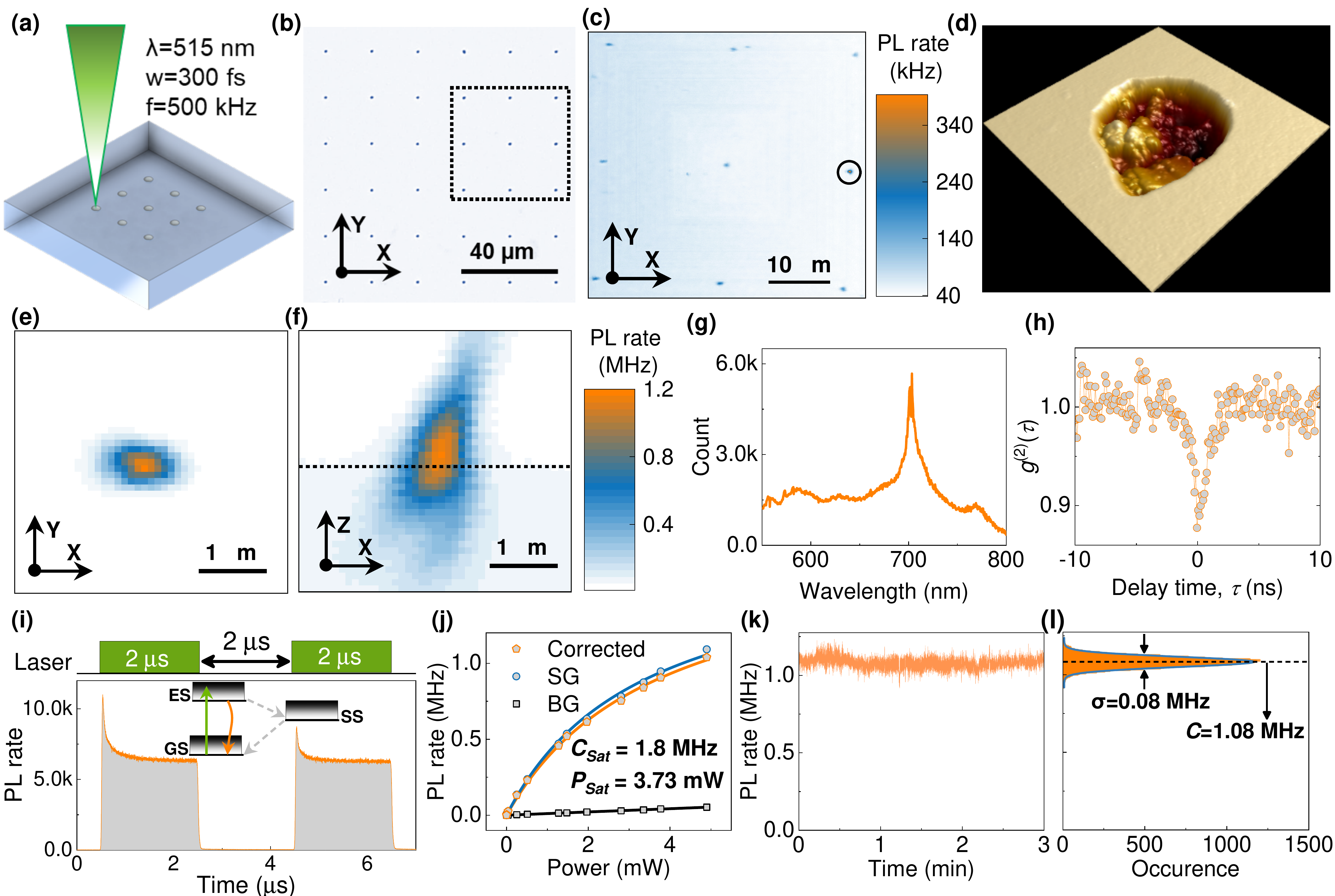}
    \caption{Laser-written antibunched emitters in GaN. (a) is a schematic diagram of laser-writing fabrication on GaN. (b) is an optical microscope image for laser-written GaN, where the area marked by a rectangular dashed box is inspected by a PL scanning of (c). (d) is the \SI{1}{\micro\metre}$\times$\SI{1}{\micro\metre} AFM image for the laser-written nano-ablation marked in (c). (e) and (f) are the xy and xz PL maps for the laser-written emitter marked in (c), where the horizontal black dashed line in (f) represents the GaN surface. (g) and (h) are the PL emission spectrum and photon emission correlation spectrum for this laser-written emitter. (i) is the time-resolved PL spectrum under double pulse laser excitation, with the insert showing a three-energy level system. (j) is the power-dependent PL saturation measurement. (k) is the PL time trace of the emitter sampled into \SI{10}{ms} bins under CW laser excitation. (l) is the corresponding histogram of the photon distribution.}
    \label{Fig.1}
\end{figure*}
The sample used in this experiment is a free-standing single-crystal nominally undoped GaN substrate with (0001) C-plane orientation. As shown in Fig. \ref{Fig.1}(a), a commercial femtosecond laser (Menlo Systems BlueCut) that produces linearly polarized pulses with a wavelength of \SI{515}{\nano\metre}, a repetition rate of 500 kHz, and a duration of \SI{300}{\femto\second} was used for femtosecond laser-writing. A 100× oil immersion objective with a numerical aperture (NA) of 0.9 is used to focus the laser beam  in GaN. The pulse energy was controlled using a combination of a motorized half-wave plate and a fixed linear polarizer. There are four different laser fabrication sets with laser exposure pulse numbers of \textit{N} = 1, 2, 5, 10, respectively. As shown in Fig. \ref{Fig.1}(b), in each set, the laser energies are spaced with \SI{20}{\micro\metre} from \textit{E} = 5-510 nJ, where there are 5 trials for each laser-writing parameter.  

\subsection{\label{Laser-written bright antibunched emitters}Laser-written bright antibunched emitters}

 The AFM scanning image in Fig. \ref{Fig.1}(d) shows that the femtosecond laser produces a nano-ablation with a rough interior. The AFM scanning measurement details are in the supplementary material (SM). Our hypothesis is that high-energy femtosecond laser results in strong non-linear absorption in GaN\cite{Eaton2019QuantumIrradiation}, which breaks the GaN lattice bond and leaves a nano-ablation \cite{10.1063/5.0160067}. After annealing, there is no emission in the laser-written areas where the fabrication laser energy is below the ablation threshold. The detailed annealing process is discussed in \ref{Annealing study}.  In contrast, an array of emitters is observed around the laser-written region where the laser energy is over the threshold, as shown in Fig. \ref{Fig.1}(c). The xy PL map of Fig. \ref{Fig.1}(e) and the xz PL map of Fig. \ref{Fig.1}(f) reveal a \SI{}{MHz}-rate point-like  emission at this location. The PL emission spectrum in Fig. \ref{Fig.1}(g) shows a sharp peak around \SI{702}{nm}, which is inconsistent with broadband emission in previous studies\cite{Saleem:18,Castelletto2020ColorSemiconductors}. PECS data, recorded by two detectors in a Hanbury-Brown and Twiss interferometer, in Fig. \ref{Fig.1}(h) shows antibunching behavior with $g^{(2)}$(0)<0.9, consistent with the creation of multiple quantum emitters within the nano-ablation. 

To understand these emitters' bunching behavior and energy structure\cite{Guo2024EmissionEmitters}, a TRPL spectrum is recorded under two \SI{2}{\micro\second} duration laser pulse excitation with a \SI{2}{\micro\second} space, as shown in Fig. \ref{Fig.1}(i). The spacing between each double-pulse train is \SI{50}{\micro\second} to allow the ground state population to reset. From the shelving behavior during the laser pulse excitation,  we deduce that at least three energy levels are present in these emitters, which include the ground state (GS), excited state (ES), and shelving state (SS), consistent with the previous study\cite{PhysRevB.97.165202}. In Figs. \ref{Fig.1}(j), the excited power-dependent PL intensity is fitted by 
\begin{equation}
C(P)= \frac{{C_{\mathrm{sat}}}P}{{P+P_{\mathrm{sat}}}} \
\label{eq:Psat}
\end{equation}
where $C(P)$ is the steady-state PL rate as a function of power \emph{P}, $C_{\mathrm{sat}}$ is the saturation PL rate, and $P_{\mathrm{sat}}$ is the corresponding saturation power. A saturation PL rate of \SI{1.8}{MHz} is achieved with a \SI{3.73}{mW} of saturation power. We also recorded the time trace of this emitter's PL rate for 3 minutes, where the emission shows a mean count rate of \SI{1.08}{MHz}, and displays a variance of \SI{0.08}{MHz}.

\subsection{\label{Laser-writing parameters study for laser-written nano-ablations with quantum emitters}Laser-writing parameters for nano-ablations with quantum emitters}

\begin{figure*}[ht]
\includegraphics[width=\textwidth]{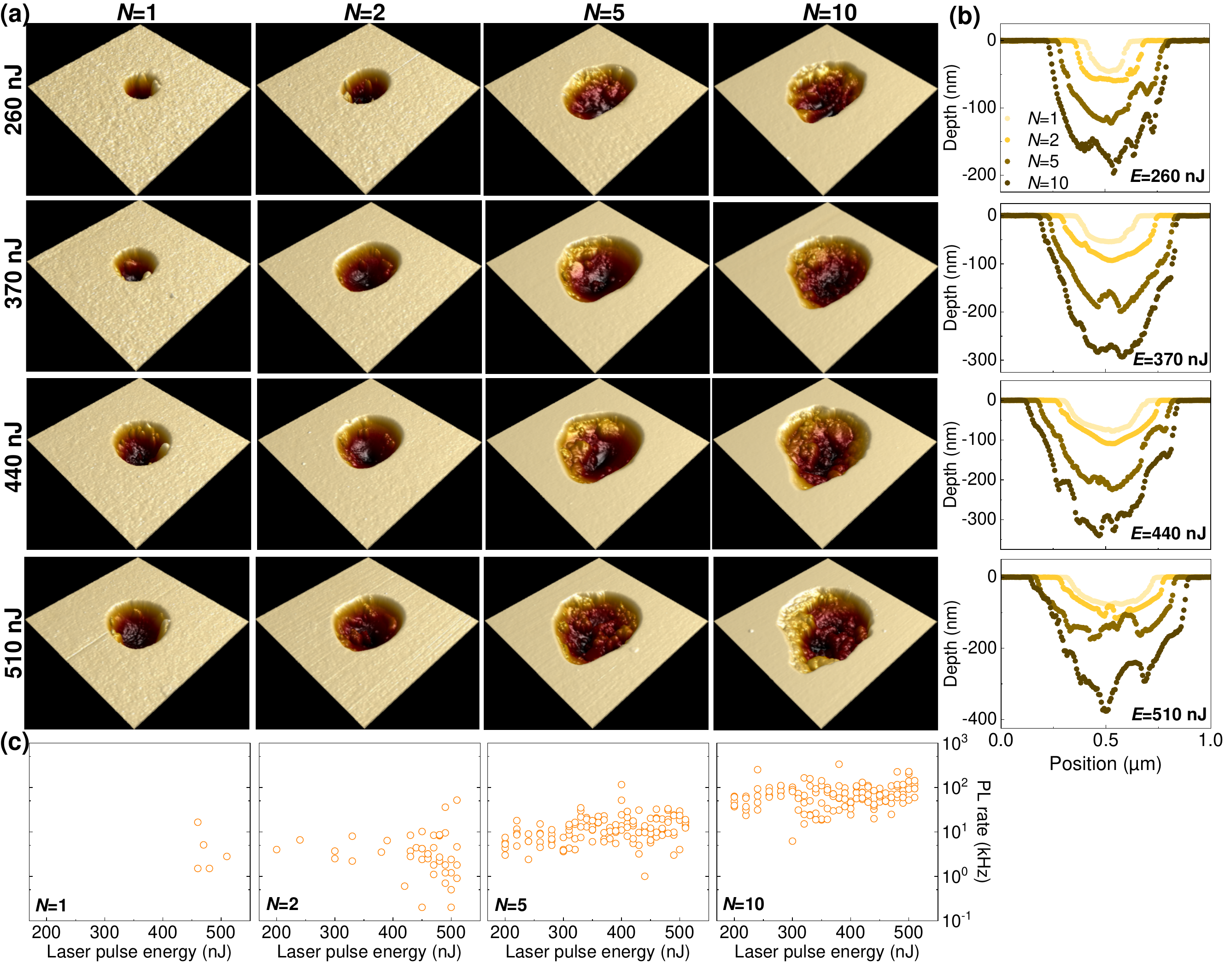}
\caption{AFM images and yield analysis for the emitters in laser-written nano-ablations in GaN. (a) summarizes the \SI{1}{\micro\metre}$\times$\SI{1}{\micro\metre} AFM images of the laser-written nano-ablations with different energies (rows) and pulse numbers (columns). (b) shows the cross sections from AFM images for different laser pulse numbers and different laser energies (columns). (c) displays the PL emission of laser-written emitters in different pulse numbers as a function of laser-writing energies.}
\label{Fig.2}
\centering
\end{figure*}

A statistical study of these laser-written nano-ablations is performed by comparing their AFM images and PL emission. First of all, from their AFM images in Fig. \ref{Fig.2}(a) and cross-section in Fig. \ref{Fig.2}(b), both increased laser energies and pulse numbers result in increased depth and diameter of the laser-written nano-ablations. Secondly, compared to laser energies, laser pulse numbers have a strong impact on the roughness of these nano-ablations. Specifically, the nano-ablations written by laser pulse number $N$=1, 2 exhibit smooth interface, while the nano-ablations fabricated by laser pulse number $N$=5, 10 feature relatively rough morphology, as shown in the cross-sections displayed in Fig. \ref{Fig.2}(b). Thirdly, Fig. \ref{Fig.2}(c) shows the yield of emitters in the nano-ablations increased as laser pulse numbers increased. Less than 4\% of the \textit{N}=1 laser-written nano-ablations show resolved PL emission. For \textit{N}=2, we find 33\% of nano-ablations display emission, and for \textit{N}=5 and \textit{N}=10 over 97\%. Finally, we note that laser energies show a weak relation to the PL rate of laser-written emitters. These observations lead us to conclude that the increased laser pulse number results in complicated surface structures of these nano-ablations which provide the emissive sources and/or the structural defect which leads to the emission. Additionally, a single shot of femtosecond laser creates a laser-written nano-ablation with a smoother surface. At single pulse laser energies less than \SI{260}{nJ}, the nano-ablation's diameter and depth will be reduced to $\sim$\SI{100}{\nano\metre} and $\sim$\SI{45}{\nano\metre}, which is less than half of the optical diffraction limit. This precise positioning ability potentially paves the way for efficient direct laser-written nanoholes in GaN coupled to photonic structures\cite{Ruelle:19}.

\subsection{\label{Photodynamics study on laser-written emitters in GaN} Photodynamics study on laser-written emitters in GaN}

\begin{figure}[ht]
    \centering
    \includegraphics[width=8.5cm]{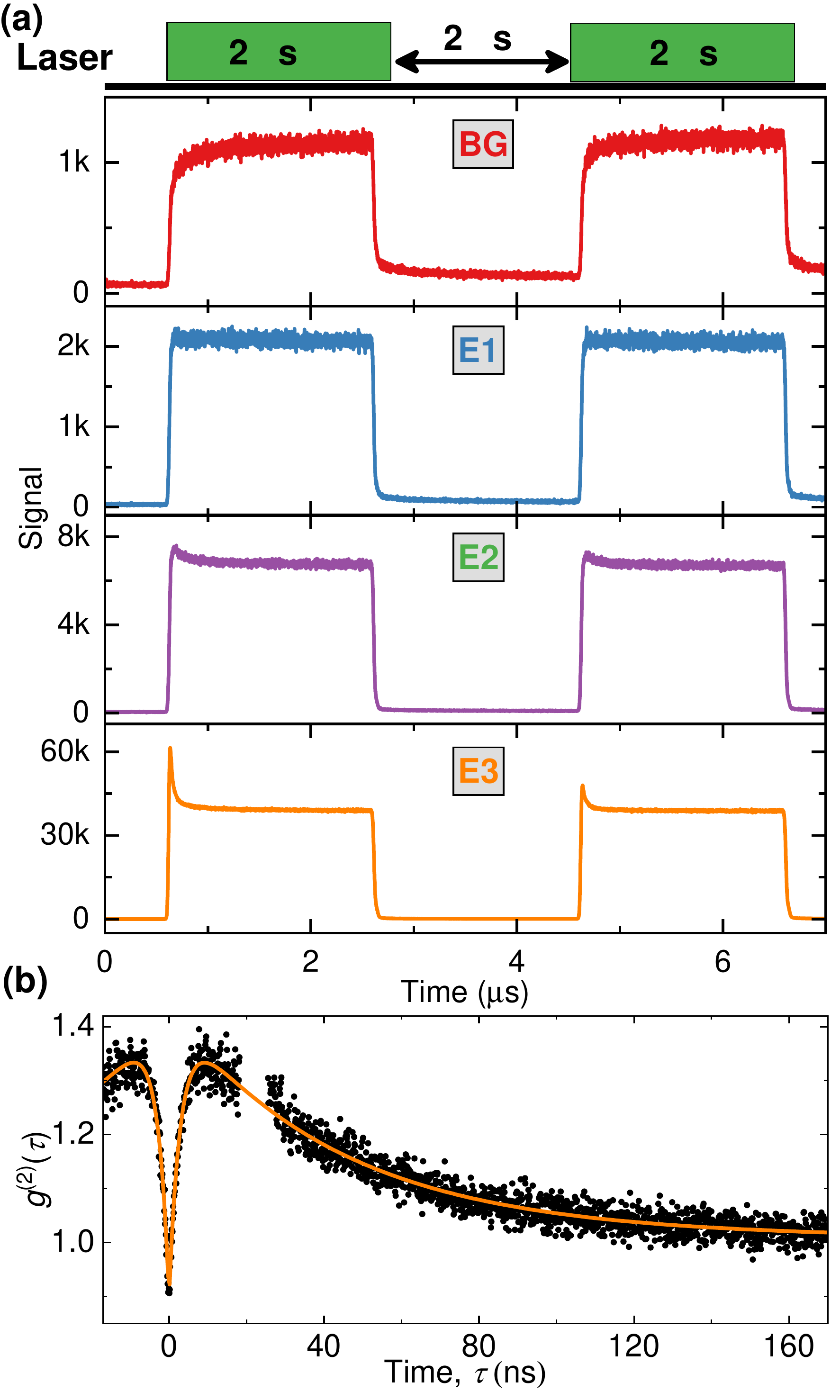}
    \caption{Photodynamics study of laser-written emitters in GaN. (a) TRPL study for different emitters and background under double pulse excitation. (b) PECS for E3}
    \label{Fig.3}
\end{figure}

We investigate the photodynamics of these laser-written emitters by TRPL and PECS. Based on the TRPL in Fig. \ref{Fig.3}(a). All emitters feature a PL peak at the start of the first laser pulse, as shown in Fig.  \ref{Fig.3}(a). In contrast, the background signal intensity linearly scales with the laser power and shows the same temporal profile as the excitation laser, corresponding to the rise of the signal from the AOM (Fig. \ref{Fig.3}(a), top panel). Emitter E1 has the weakest PL emission and does not display strong shelving decay in its TRPL. This type of behavior is mostly observed in $N$=1, 2 laser pulse written areas. Emitter E2 emits a stronger PL signal, features an additional shelving decay within hundreds of \SI{}{\nano\second} in its TRPL, and is found in laser-written regions with $N$=5,10. Emitter E3 features \SI{}{MHz} PL rate and exhibits tens of \SI{}{\nano\second} shelving process. Approximately 5\% of the nano-ablations with laser-pulse number $N$=10 are of this type. In the time between laser pulses, the system does not fully relax to the ground state in E2 or E3, resulting in a lower amplitude intensity at the start of the second pulse. This is indicative of a several \SI{}{\micro \second} shelving state lifetime\cite{Guo2024EmissionEmitters}.

Less intense emitters like E1 and E2 did not exhibit antibunching or bunching in their PECS. In contrast, E3 shows antibunching signal in its PECS in Fig. \ref{Fig.3}(b). The $g^{(2)}$(\(\tau\)) data is fitted using the empirical equation,
\begin{equation}
\begin{split}
g^{(2)}(\tau)= & 1-C_1e^{-|\tau|/\tau_1}
+C_2e^{-|\tau|/ \tau_2}+C_3e^{-|\tau| /\tau_3} \
\label{eq:g2_empirical}
\end{split}
\end{equation}
Here, $\tau_{1}$ is the antibunching time, $C_{1}$ is the antibunching amplitude, $\tau_{i}$ for \emph{i} \(\geq\) 2 are bunching times, and $C_{i}$ for \emph{i} \(\geq\) 2 are the corresponding bunching amplitudes. The fitting parameters are shown in table \ref{tb2:g2fittingparameters}.

\begin{table}[ht]
  \caption{Fitting parameters of $g^{(2)}(\tau)$}
    \label{tb2:g2fittingparameters}
  \begin{tabular}{lll}
    \hline
   Parameters&  Value & Standard Error
\\
 $C_1$& 0.54& 0.005\\
 $\tau_1$ (ns)& 2.90& 0.06\\
 $C_2$& 0.42& 0.003\\
 $\tau_2$ (ns)& 42.93& 0.59\\
 $C_3$& 0.019& 0.002\\
 $\tau_3$ (ns)& 290.84& 56.46\\\hline
 \end{tabular}

\end{table}

Overall, the PL emission of QEs is defined by its excited state lifetime and shelving process\cite{Guo2024EmissionEmitters,Patel2022ProbingNitride}. For TRPL measurement, the time trace is partly determined by the population in the ground state at the beginning of the laser pulse. Therefore, E3 exhibits a sharp rise at the beginning of laser pulses due to the laser-pumped occupation of its excited state and subsequently follows an exponential decay until a steady state is reached due to the shelving process. This also results in the bunching behavior of E3 in Fig. \ref{Fig.3}(b). We conclude that the shelving process prevails in laser-written emitters, governing the dynamics of emission from their atomic-like energy levels. 

\subsection{\label{Annealing study}Annealing study}

\begin{figure}[ht]
    \centering
    \includegraphics[width=8.5cm]{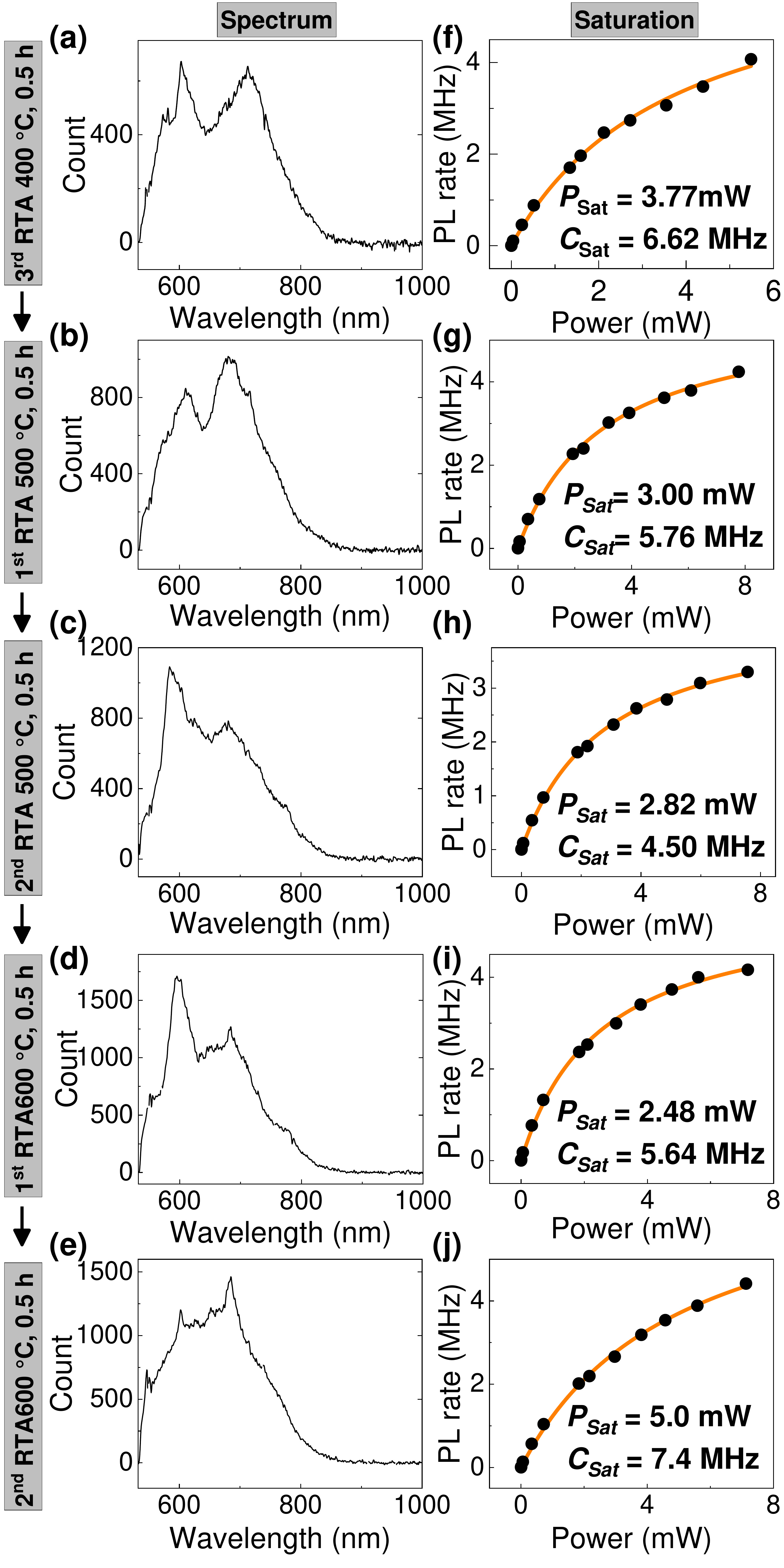}
    \caption{The spectra (a-e) and power-dependent PL saturation behaviors (f-j) for E3 during the RTA annealing.}
    \label{Fig.4}
\end{figure}
Before annealing, the laser-written ablations are difficult to resolve from the PL scanning as shown in Fig. \ref{Fig.S1}(a). Only a few laser-written ablations emit weak PL emission above the background level. These all show dipole-like excitation polarization dependence and a broadband PL spectrum from \SI{550}{\nano\metre} to \SI{800}{\nano\metre} without the antibunching (Fig. \ref{Fig.S1}). Thus, these laser-written regions might be some optically active structural dislocation that cannot form a stable QE without further annealing, or there may be many emitters within the confocal laser spot. Subsequently, a series of rapid thermal annealing (RTA) processes are used to activate the laser-written emitters in nitrogen gas ambiance. After the first 0.5 h of 400 °C annealing, the laser-written areas become brighter and show weak dependence on laser-written energies, with an improved PL contrast compared to the background as shown in Fig. \ref{Fig.S2}. Most of the laser-written emitters show similar properties as the emitters before annealing, exhibiting dipolar-like excitation polarization, a broadband spectrum from 550-\SI{800}{\nano\metre}. Some blinking emitters are also found on the edge of laser-written nano-ablation, as circled in Fig. \ref{Fig.S2}(a), and close to the surface of the substrate, as shown in Fig. \ref{Fig.S2}(b). More importantly, they show antibunching in their PECS. However, these QEs are highly unstable and suffer from photo-bleaching, which makes them unable to collect their spectra. 

After the second 0.5 h of 400 °C annealing, an increasing number of blinking QEs are observed in the laser-written region, some with weak antibunching (g$^{(2)}$(0) $\approx$ 0.9). Their emission is broad between 550-800 nm but with a few sharp peaks. Their power-dependent PL shows the saturation behavior. After the third 0.5 h of 400 °C annealing, we found some stable and bright emitters with \SI{}{MHz} of PL rate around the laser-written region, such as E3 in Fig. \ref{Fig.3}.

We monitor E3 over several further cycles of annealing, showing substantial changes to saturation rate and spectrum in Fig. \ref{Fig.4}. There are two peaks around \SI{600}{\nano\metre} and \SI{700}{\nano\metre}, whose intensities are affected via annealing. This highlights the dynamic processes of creation and annihilation for different species of QEs in laser-written spots. After three anneals of 0.5 h at 400 °C, two anneals 0.5 h of 500 °C, and one anneal of 0.5 h 600 °C, a sharper peak is resolved at a wavelength of \SI{600}{\nano\metre}. However, after the second 0.5 h of 600 °C, these sharp peaks are suppressed by the strong phonon side band, exhibiting the unstable PL emission with photobleaching. 

Another signature of a quantized emitter is its photoluminescence (PL) intensity saturation with excitation. From Figs. \ref{Fig.4}(f-j), the $P_{sat}$ decreases with the annealing process until the second time 0.5 h of 600 °C annealing, and then increases again as new features appear in the spectrum for subsequent anneals. These results suggest that 0.5 h of 600 °C RTA annealing is a threshold for the creation of stable quantum emitters. This may be related to a previous annealing study\cite{10.1063/5.0187072}, which predicted diffusion of nitrogen-vacancy centers occurs at above 500 °C.

\section{\label{Conclusion and outlook} Conclusion and outlook}
 
We report the engineering of a regularly-spaced array of QEs in GaN via femtosecond laser-writing and subsequent annealing. Bright, antibunched emission with \SI{}{\mega\hertz} PL rates and sharp spectral peaks are deterministically created in the laser-written nano-ablations. The laser-writing effect is also investigated as a function of laser energy, pulse number and annealing using AFM and PL studies. We find laser-written nano-ablations with an increased laser pulse number ($N$=5, 10) annealed at 500-600 °C are optimal to create emitters that display the signatures of quantized electronic states. Our study paves a promising way to scalable engineering of QEs inside photonic nanostructures and integrated quantum circuits. Future works should focus on increasing the yield of bright laser-written emitters and fabricating the emitters at a single level, possibly by using in-situ monitoring of emitter creation during the laser writing process. Moreover, our study also demonstrates an efficient direct laser nano-ablation process for GaN, providing important information for the fabrication of laser-written photonic circuits.

\begin{acknowledgments}
The authors acknowledge financial support provided by EPSRC via Grant No. EP/T017813/1 and EP/03982X/1 and the European Union's H2020 Marie Curie ITN project LasIonDef (GA No. 956387) and  Sêr Cymru National Research Network in Advanced Engineering and Materials. IFN-CNR is thankful for support from the projects QuantDia (FISR2019-05178) and PNRR PE0000023 NQSTI funded by MUR (Ministero dell'Università e della Ricerca). Device processing was carried out in the cleanroom of the ERDF-funded Institute for Compound Semiconductors (ICS) at Cardiff University. We wish to thank ICS staff and Miguel Alvarez Perez for technical support.
\end{acknowledgments}

\section*{Author Contributions}
The corresponding author identified the following author contributions, using the CRediT Contributor Roles Taxonomy standard:

YG: Conceptualization, Methodology, Software, Data curation, Formal Analysis, Investigation, Resources, Writing - Original Draft, Writing - Review \& Editing. GC: Investigation, Resources, Writing - Review \& Editing. VB: Investigation, Resources, Writing - Review \& Editing. RY: Investigation, Resources, Writing - Review \& Editing. KE: Investigation, Writing - Review \& Editing. JPH: Conceptualization, Resources, Writing - Review \& Editing, Supervision, Funding Acquisition. AJB: Conceptualization, Resources, Writing - Review \& Editing, Supervision, Funding Acquisition. SME: Conceptualization, Resources, Writing - Review \& Editing, Supervision, Funding Acquisition.

\section*{Data Availability Statement}
Data supporting the findings of this study are available in the Cardiff University Research Portal at http://doi.org/xx.xxxx
\nocite{*}
\section*{Conflict of Interest}
The authors declare no conflicts of interest.

\bibliography{References}
\clearpage
\newpage
\onecolumngrid
\begin{center}
\textbf{\large Supplemental Materials: Femtosecond laser-written nano-ablations containing bright antibunched emitters on gallium nitride}
\end{center}

\setcounter{equation}{0}
\setcounter{figure}{0}
\setcounter{table}{0}
\setcounter{page}{1}
\makeatletter
\renewcommand{\theequation}{S\arabic{equation}}
\renewcommand{\thefigure}{S\arabic{figure}}
\renewcommand{\bibnumfmt}[1]{[S#1]}
\setcounter{section}{0}

\section{\label{Confocal setup} Confocal setup}

A home-built room-temperature confocal setup is used to study the PL emission for the laser-written GaN sample. A CW \SI{532}{\nano m} crystal laser was modulated by an acoustic-optic modulator (ISOMET 553F-2) with < \SI{10}{\nano s} rise and fall time. TRPL was binned with 1 ns resolution. A 2-axis Galvo mirror (GVS002) and 100$\times$ Nikon objective with NA=0.9 were integrated into a 4f imaging system for 2D x-y scanning. Depth scanning (z) was implemented by a closed-loop piezo sample stage. The PL was optically filtered by the dichroic mirror, \SI{532}{\nano m} long-pass filter before detection on SPCM-AQRH silicon avalanche photodiodes (Excelitas) or a spectrometer with a silicon CCD. The optional ND filter is also used to keep the PL rate within the APD's linear response range (\SI{2}{\mega\hertz}).

\section{\label{AFM measurement}AFM measurement}

The AFM measurement was conducted in commercial Bruker AFM microscopes. The scanning tip is the silicon tip on the nitride lever. The scan parameters are \SI{1}{Hz} of scanning rate with 256 samples and $0\degree$ of scan angle. The AFM image was analysed and plotted by the Brucker commercial software (Nanoscope Analysis 3.0). 

\section{\label{PL study for laser-written GaN before annealing} PL study for laser-written GaN before annealing}

\begin{figure*}[ht]
\includegraphics[width=0.73\textwidth]{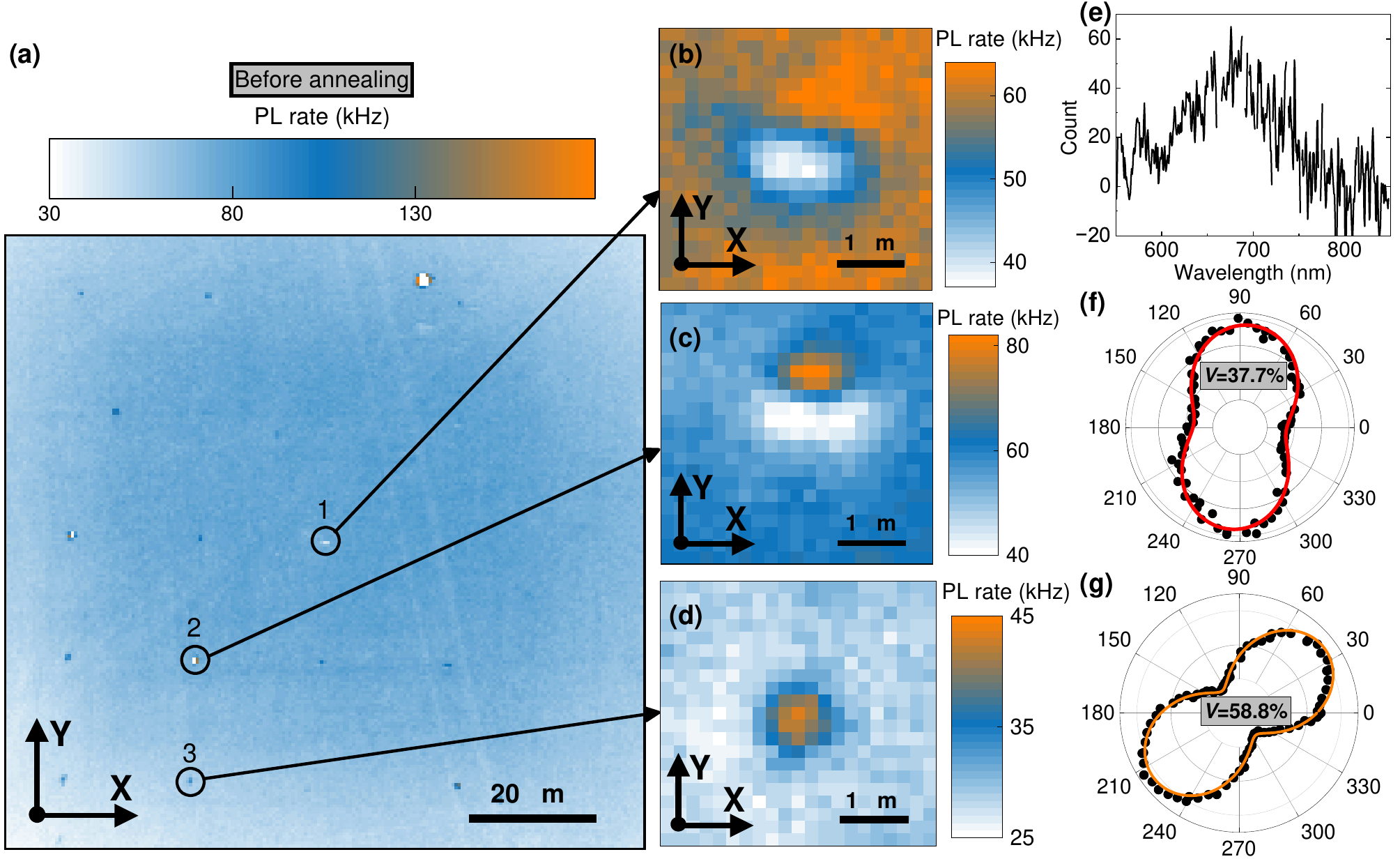}
\caption{The PL study of laser-written GaN before any annealing process. (a) PL maps of laser-written GaN.(b-d) are the high-resolution PL maps of the laser-written areas labeled as 1-3, respectively in (a). (e) represents the spectrum of emitters in (d). (f) and (g) are the excitation polarization plots for the emitters in (c) and (d), respectively.}
\label{Fig.S1}
\centering
\end{figure*}
As shown in Fig. \ref{Fig.S1}(a), the regular emitter grids have been observed from the laser-written area in the GaN substrate. The PL emission of these laser-written spots is weak relative to the background with a small dependence on the laser energy, consistent with previous studies\cite{Castelletto2020ColorSemiconductors}. Some laser-written spots show a low-intensity center in the fabrication region as shown in Fig. \ref{Fig.S1}(b). Some display a low-intensity center with a nearby bright emitter as shown in Fig. \ref{Fig.4}(c).  These low-intensity centers are due to the laser-writing produced structural deterioration of the GaN crystal. Some laser-written areas only feature a bright emitter. The emitters in Figs. \ref{Fig.S1}(c) or (d) exhibit a broad emission from 550-\SI{800}{\nano\metre} without a clear ZPL, and optical dipolar-like excitation polarization as shown in Fig. \ref{Fig.S1}(f) and (g). The excitation polarization plots are fitted by
\begin{equation}
C(\theta)=A\cos^2{(\theta-\theta_0)}+B
\label{eq:Abspolar}
\end{equation}
where  \textit{A} is the amplitude, $\theta_0$ is the polarization angle of the maximum PL rate, and \textit{B} is the offset\cite{Patel2022ProbingNitride}. Their visibility \textit{V} of 37.7\% and 58.8\% for emitters in Fig. \ref{Fig.S1}(c) and (d) are calculated by the equation\cite{Patel2022ProbingNitride},
\begin{equation}
    V=\frac{C\textsubscript{Max}-C\textsubscript{Min}}{C\textsubscript{Max}+C\textsubscript{Min}}=\frac{A}{A+2B}
\label{Vis}
\end{equation}
where $C\textsubscript{Max}$ and $C\textsubscript{Min}$ are the maximum and minimum intensities, respectively.  

However, no antibunching or bunching behaviors were found in their PECS. And their PL intensity scales linearly with laser power.

\section{PL study for laser-written G\lowercase{a}N after first 0.5 h of 400 °C annealing}

\begin{figure}[ht]
\includegraphics[width=10cm]{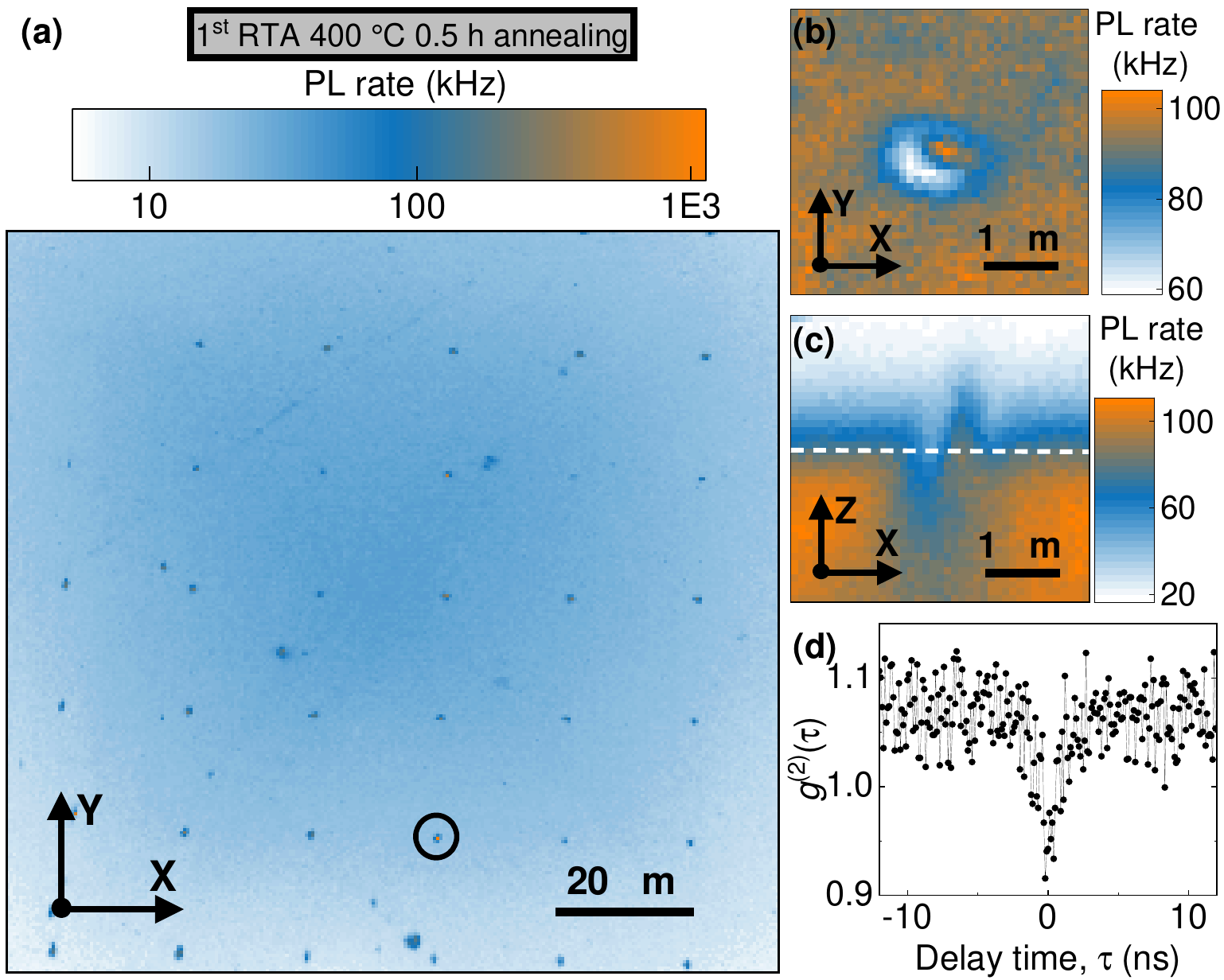}
\caption{{The PL study of laser-written GaN after the first 0.5 h of 400 °C. (a) PL maps for laser-written GaN. (b) and (c) are the x-y and x-z PL maps of the laser-written grids marked in (a), where the white dashed line represents the GaN substrate surface. (d) is the PECS of the QEs in (a-c).}}
\label{Fig.S2}
\centering
\end{figure}

\end{document}